\documentstyle[12pt]{article}
\newcommand{\bea}{\begin{eqnarray}}
\newcommand{\eea}{\end{eqnarray}}
\newcommand{\beq}{\begin{equation}}
\newcommand{\eeq}{\end{equation}}
\newcommand{\nn}{\nonumber}

\newcommand{\dst}{\displaystyle}

\begin{document}
\begin{center} {\Large \bf
Fermi coordinates for weak gravitational fields }
\footnote{To appear in Phys.~Rev.~D}\\[1cm]
Karl-Peter Marzlin \\[2mm]
Fakult\"at f\"ur Physik 
der Universit\"at Konstanz\\
Postfach 5560 M 674\\
D-78434 Konstanz, Germany
\footnote{e-mail: peter@spock.physik.uni-konstanz.de}
$ $ \\[4mm]
\begin{minipage}{15cm}
\begin{abstract}
We derive the Fermi coordinate system of an observer in
arbitrary motion in an arbitrary weak gravitational field valid
to all orders in the geodesic distance from the worldline
of the observer. In flat space-time this
leads to a generalization of Rindler space for arbitrary
acceleration and rotation. The general approach is applied
to the special case of an observer resting with respect to
the weak gravitational field of a static mass distribution.
This allows to make the correspondence between general
relativity and Newtonian gravity more precise.
\end{abstract} \end{minipage}\end{center}
$ $ \\[3mm]
\section{Introduction}
General relativistic studies of physical situations far away
from black holes, neutron stars, or the big bang are often
based on the linearization of Einstein's field equations and
the assumption that the gravitational field is weak enough
to allow a perturbational approach. The metric
of space-time is written in the form
\beq g_{\mu \nu} = \eta_{\mu \nu} + h_{\mu \nu} \; ,\; |h_{\mu
    \nu}| \ll 1 \eeq
where $\eta_{\mu \nu}$ is the Minkowski metric and $h_{\mu \nu}$
are small deviations which are treated only to {\em first order
wherever they occur}. Although this form of the metric restricts
the metric to be nearly Minkowskian one has a remaining
freedom to choose a coordinate system. For instance,
to calculate gravitational waves as small
perturbations in the vacuum one usually uses the freedom of
coordinate (or gauge) transformations to impose the restriction
$h_{\mu \; \; \; , \nu}^{\; \, \nu} - {1\over 2}
 h^\alpha_{\; \alpha, \mu} =0 $ (harmonic gauge)
on the gravitational field thus
fixing to a certain amount the coordinate system (see, e.g.,
Ref. \cite{MTW}). As discussed by Faraoni \cite{faraoni92} this
choice of coordinates is very convenient for the study of
gravitational waves, but it is difficult to describe the
motion of the observer or the detector in this case.

The study of the observer is best done in his Fermi coordinate
system \cite{manasse63,MTW}. This is in some sense
a generalization of the
notion of a frame of reference for an inertial observer in
flat space to curved space-time. In these coordinates the metric
tensor itself contains at least in the approximations
which have been examined
only quantities which are invariant under coordinate
transformations like the proper time of the observer,
geodesic distances from the worldline, and components of
tensors with respect to a tetrad.

For gravitational waves and an observer in geodesic motion
the transformation to Fermi coordinates
was done by Fortini and Gualdi \cite{fortini82}.
Because actual observers are often accelerated, and because
the applications of the weak field limit of general relativity
is not restricted to gravitational waves, it is of interest to
generalize their results to the case of an observer in arbitrary
motion in an arbitrary weak gravitational field. It is the
purpose of this letter to give this generalization.
To incorporate the acceleration and the rotation of the observer
we follow the approach of Ni and Zimmermann \cite{ni78}.

After the performance of the general construction in Sec.~2
we apply the result to an observer resting with respect to
an arbitrary mass distribution. This physical situation is
mostly studied in the context of Newtonian gravity, and it
is often used to demonstrate that the Newtonian picture is a
certain limit of general relativity. While the last claim
it is certainly true it was never studied in the context of
Fermi coordinates, i.e., in some kind of reference system of
the observer. We will give the corresponding analysis in
Sec.~3. In Sec.~4 the results are reviewed and some comments
are made to the existing work on gravitational waves.
Our metric conventions are that of Ref. \cite{MTW}, i.e. the
signature of the metric is +2. Greek indices run from 0 to 3,
latin indices from 1 to 3. Tetrad indices are underlined.
We use units with c=1.
\section{Derivation of the general result}
The observer is assumed to move (in the coordinate system
$y^\mu$) on the worldline $z^\mu (\tau )$
governed by the equation
\beq \ddot{z}^\mu + \Gamma^\mu_{\; \nu \lambda} \dot{z}^\nu
     \dot{z}^\lambda = a^\mu \label{geodamu} \eeq
where $\tau$ is the proper time of the observer and $a^\mu$ is
the four-acceleration. A dot denotes
the derivative with respect to $\tau$. To construct the
hypersurfaces of constant $\tau$ we consider the family of
geodesics $y^{\mu} (\tau ,s)$ with geodesic length $s$
starting at $y^\mu (\tau ,0) = z^\mu (\tau )$  which fulfill the
equation
\beq y^{\prime \prime \mu} + \Gamma^\mu_{\; \nu \lambda}
     y^{\prime \nu} y^{\prime \lambda} = 0 \label{geodgl}\eeq
where a prime denotes the derivative with respect to $s$.
Their tangent vector on the worldline has to be
perpendicular to the tangent vector of the worldline $\dot{z}
^\mu (\tau )$, that is
\beq \dot{z}^\mu y^\prime_{\mu}|_{s=0} =0 \; . \label{perp}
     \eeq
Decompose $y^\mu (\tau ,s)$ into a Minkowski part $y^\mu_M$ and
a part $y^\mu_h$ which is of the order of $h_{\mu \nu}$. The
same can be done with the worldline $z^\mu$ of the observer.
The Minkowski part of the solution of Eq. (\ref{geodgl}) is
easily found by noting that in flat space all geodesics are
straight lines. Thus
\beq y^\mu_M (\tau ,s)  = z^\mu_M (\tau ) + s v^\mu_M(\tau )\eeq
where $v^\mu_M$ is some vector which is perpendicular to
to $\dot{z}^\mu_M$ in the Minkowskian
sense, i.e. $v^\mu_M \dot{z}
^\nu_M \eta_{\mu \nu} = 0$. Inserting this into Eq.
(\ref{geodgl}) leads to
\beq y^{\prime \prime \mu}_h + \eta^{\mu \rho} (h_{\rho \nu ,
     \lambda} - \frac{1}{2} h_{\nu \lambda ,\rho}) v^\nu_M
     v^\lambda_M = 0 \label{hgeod} \eeq
Up to now each factor of $h_{\mu \nu}$
was taken at the point $y^\mu = y^\mu_M + y^\mu_h$. By making
a Taylor expansion around $y^\mu_M$ one can see that all
terms including derivatives of $h_{\mu \nu}$ are of higher order
in $h_{\mu \nu}$ so that one can restrict the sum to the lowest
order,
\beq h_{\mu \nu}(y^\rho_M + y^\rho_h ) = h_{\mu \nu} (y^\rho_M)
     + O((h_{\mu \nu})^2) \; . \eeq
The general solution of Eq. (\ref{hgeod}) is then given by
\beq y^\mu_h (\tau ,s) = C_2^\mu (\tau ) + s C_1^\mu (\tau ) -
     \int_0^s v^\nu_M
     h^\mu_{\; \nu} (y^\mu_M (\tau ,s^\prime )) ds^\prime +
     \frac{1}{2} \eta^{\mu \rho} v^\nu_M v^\lambda_M
     \int_0^s ds^\prime \int_0^{s^\prime} ds^{\prime \prime}
      h_{\nu \lambda ,\rho}(y^\mu_M (\tau ,s^{\prime \prime}))
      \; . \eeq
The four-vectors $C_i^\mu$ have to be determined by
the condition (\ref{perp})  together with $y^\mu (\tau ,0) =
z^\mu (\tau )$ and $y^{\prime \mu} y^\prime_{\mu}|_{s=0} = 1$.
To make contact with Fermi coordinates and to give the result
in a convenient form we first introduce an orthonormal tetrad
$e^\mu_{\underline{\alpha}}(\tau )$
defined in the tangent space of $z^\mu (\tau )$
which fulfills
$e^\mu_{\underline{0}} = \dot{z}^\mu (\tau )$ and has the
equation of motion \cite{MTW}
\beq \frac{\dst D e_{\underline{\alpha}}}{d \tau} = -
     \hat{\Omega}\cdot e_{\underline{\alpha}}
     \label{tetdgl}\eeq
with
\beq \hat{\Omega}^{\mu \nu} = a^{\mu} \dot{z}^{\nu} - a^{\nu}
     \dot{z}^{\mu} + \dot{z}_{\alpha} \omega_{\beta}
     \varepsilon^{\alpha \beta \mu \nu} \label{tetom} \eeq
where $\omega^\alpha$ is the four-rotation of the tetrad.
If the condition (\ref{perp}) is fulfilled we can write
\beq y^{\prime \mu} (\tau ,0) = v^\mu_M + C_1^\mu  - v^\nu_M
     h^\mu_{\; \nu} (\tau ) = \alpha^{\underline{i}}
     e^\mu_{\underline{i}} \; . \eeq
The parameters $\alpha^{\underline{i}}$ determine the direction
of the geodesic at the worldline and $h_{\mu \nu} (\tau )$ is a
shorthand for $h_{\mu \nu} (y^\mu (\tau ,0))$. Taking all
together the family of geodesics perpendicular to the tangent
vector of the worldline is parametrized by
$\alpha^{\underline{i}}$ and is given by
\beq y^\mu (\tau ,s) = z^\mu (\tau )+ s \alpha^{\underline{i}}
     (e_{\underline{i}}^\mu + h_{\underline{i}}^{\; \mu} (\tau )
     ) - \alpha^{\underline{i}} \int_0^s h_{\underline{i}}^{\;
     \mu} ds^\prime + \frac{1}{2} \eta^{\mu \rho}
     \alpha^{\underline{i}} \alpha^{\underline{j}} \int_0^s
     ds^\prime \int_0^{s^\prime} ds^{\prime \prime}
     h_{\underline{ij},\rho} \; . \label{geodaete}\eeq
Here and in the remainder we use the transformation of
space-time indices to tetrad indices, e.g. $X_{\underline{
\alpha}}= X_{\mu} e^\mu_{\underline{\alpha}}$. We use
this typing also if the index is a derivative.
It is now convenient to make a Taylor expansion of the integrals
in Eq. (\ref{geodaete}). It is obvious that
\beq \left ( \frac{d}{ds} \right )^n h^\mu_{\;
     \underline{i}} (y^\mu_M = z^\mu_M + s v^\mu_M) =
     h^\mu_{\; \underline{i},\underline{k}_1 \cdots
     \underline{k}_n} (y^\mu_M )\,  \alpha^{\underline{k}_1}
     \cdots \alpha^{\underline{k}_n} \eeq
holds. We introduce the Fermi coordinate system
$x^\mu$ in a weak gravitational field by setting $x^0 = \tau$
and $x^i = s \alpha^{\underline{i}}$ as proposed by
Manasse and Misner \cite{manasse63}. The transformation from the
coordinate system $y^\mu$ to Fermi coordinates is then given
by Eq. (\ref{geodaete}), or, after the Taylor expansion, by
\bea y^\mu (x^\mu ) &=& z^\mu (x^0) + x^i (e^\mu_{\;
     \underline{i}}(x^0) + h^\mu_{\; \underline{i}} (x^0) ) -
     \sum_{l=0}^{\infty} \frac{1}{(l+1)!}\, h^\mu_{\;
     \underline{
     m},\underline{k}_1 \cdots \underline{k}_l}(x^0)\, x^m
     x^{k_1} \cdots x^{k_l} \nn \\
     & & + \frac{1}{2} \sum_{l=0}^{\infty} \frac{1}{(l+2)!}\;
     h_{\underline{m n}\; \; ,\underline{k}_1 \cdots
     \underline{k}_l}^{\; \; \; \; \; \, , \mu}(x^0)\, x^m x^n
     x^{k_1} \cdots x^{k_l} \; . \eea
With the aid of this formula it is straightforward to derive
the metric in Fermi coordinates by using Eq. (\ref{tetdgl}) and
\beq g_{\alpha^\prime \beta^\prime}(x^\mu ) = \frac{\partial
     y^\mu}{\partial x^{\alpha^\prime}} \frac{\partial y^\nu}{
     \partial x^{\beta^\prime}} g_{\mu \nu}(y^\mu (x^\mu ))
     \; . \eeq
Note that the factor of $g_{\mu \nu}$ on the r.h.s. must also
be expanded in terms of $x^i$.
The result is
\bea g_{00} &=& - (1+ a_{\underline{i}} x^i)^2 + (\vec{\omega}
     \times \vec{x})^2 - \gamma_{\underline{00}} - 2 (\vec{
     \omega}\times \vec{x})_{\underline{i}} \gamma_{\underline{
     0i}} -(\vec{\omega}\times \vec{x})_{\underline{i}}
     (\vec{\omega}\times \vec{x})_{\underline{j}} \gamma_{
     \underline{ij}} \nn \\
     g_{0i} &=& (\vec{\omega}\times \vec{x})_{\underline{i}} -
     \gamma_{\underline{0i}} - (\vec{\omega}\times
     \vec{x})_{\underline{j}} \gamma_{\underline{ij}} \nn \\
     g_{ij} &=& \delta_{ij} - \gamma_{\underline{ij}} \; .
     \label{fermimetrik} \eea
The expression $(\vec{\omega} \times \vec{x})_{\underline{i}}$
denotes $\varepsilon_{\underline{ijk}} \omega^{\underline{j}}
x^k$, and  the coefficients $\gamma_{\underline{\mu \nu}}$ are
found to be
\bea \gamma_{\underline{00}} &=& \sum_{r=0}^\infty \frac{2}{(r
     +3)!}\, R_{\underline{0m0n},\underline{k}_1 \cdots
     \underline{k}_r} x^m x^n x^{k_1} \cdots x^{k_r} [ (r+3)
     + 2 (r+2) x^i a_{\underline{i}} + (r+1) (x^i a_{\underline{
     i}})^2] \nn \\
     \gamma_{\underline{0i}} &=& \sum_{r=0}^\infty \frac{2}{(r
     +3)!}\, R_{\underline{0min},\underline{k}_1 \cdots
     \underline{k}_r} x^m x^n x^{k_1} \cdots x^{k_r} [ (r+2)
     + (r+1) x^i a_{\underline{i}}] \nn \\
     \gamma_{\underline{ij}} &=& \sum_{r=0}^\infty \frac{2(r+1)
     }{(r+3)!}\, R_{\underline{imjn},\underline{k}_1 \cdots
     \underline{k}_r} x^m x^n x^{k_1} \cdots x^{k_r} \; .
     \label{coeff} \eea
where the linearized Riemann tensor is given by
\beq R_{\mu \nu \alpha \beta} = {1\over 2} [ h_{\mu \beta ,
     \nu \alpha} + h_{\nu \alpha ,\mu \beta} - h_{\nu \beta
     , \mu \alpha} -h_{\mu \alpha , \nu \beta}] \eeq
Eq. (\ref{fermimetrik}) is the main result of this letter. It
agrees with Eq. (49) of Fortini and Gualdi \cite{fortini82}
for the case of a gravitational wave and with the more general
result of Li and Ni \cite{li79} in absence of any
rotation or acceleration. It is also in concordance with the
expansion to third order in the geodesic distance $s$ from the
worldline derived by Li and Ni for an accelerated
and rotating observer \cite{li79b}.

It is worth to notice that in absence of curvature, i.e. in
flat space, all $\gamma_{\underline{\mu \nu}}$ vanish and
Eq. (\ref{fermimetrik}) becomes exact, even for strong
accelerations or rotations which may depend on the proper time.
In this case Eq.~(\ref{fermimetrik}) can be considered as the
generalization
of Rindler space-time which describes an observer with
constant acceleration in two dimensions.
\section{The resting observer in the field of a static
mass distribution}
Beside gravitational waves it is of interest to study the
Fermi coordinates of an observer who is at rest with
respect to a static mass distribution. This situation
is equivalent to those usually treated in Newtonian gravity.
Since Fermi coordinates are in this context
close to the concept of an inertial system in flat space
their analysis may give the notion of the Newtonian limit
of general relativity a more accurate form.

The gravitational field is caused by a static
mass density $\rho (\vec{y}) = T_{00} (\vec{y})$. All other
components of the energy-momentum tensor $T_{\mu \nu}$ are
assumed to vanish. In the harmonic gauge the solution of the
linearized Einstein equations is then given by \cite{MTW}
\beq h_{00}(\vec{y})=h_{i(i)}(\vec{y}) 2
     G \int \frac{\rho (\vec{y}^\prime )}{|\vec{y}
     -\vec{y}^\prime |}\, d^3 y^\prime \label{statfeld2} \eeq
where $G$ is Newton's constant and the index $i$ in brackets
denotes that no summation is understood.
Obviously, the components
$h_{\mu \nu}$ are related to Newton's potential $\phi$ by
$\phi (\vec{y})=-h_{00}/2$. The curvature tensor has in this
coordinate system the components
\bea R_{0l0m} &=& \phi_{,lm} \nn \\
     R_{0lim} &=& 0 \nn \\
     R_{imjn} &=& \delta_{mn} \phi_{,ij}+\delta_{ij} \phi_{,mn}
     -\delta_{in} \phi_{,jm} - \delta_{jm} \phi_{,in} \; .
     \label{masKr}\eea
We take the observer to be fixed to the spatial origin
of the coordinate system, $z^i =0$. By the normalization
condition $\dot{z}^\mu \dot{z}_{\mu} =-1$ and Eq.
(\ref{geodamu}) his acceleration is found to be
\beq a^0 =0 \; , \quad a^i = \partial_i \phi |_0 \eeq
which is the negative of Newton's acceleration. This is
reasonable since the Newtonian acceleration describes the
apparent acceleration of freely falling objects (the apple)
as seen by the observer and is therefore the negative of the
actual acceleration of the observer. An appropriate tetrad is
\beq e_{\underline{\alpha}}^\mu = \delta_{\alpha}^\mu +
     O(h_{\mu \nu})\; . \label{TetWeak}\eeq
We can neglect the first order part since we will be
concerned with tensors which are already of first order in
$h_{\mu \nu}$.
It follows that the components of the curvature tensor and
the acceleration with respect to the tetrad are the same as
those taken with respect to the coordinates $y^\mu$.

We now turn to the calculation of the quantities
$\gamma_{\underline{\mu \nu}}$. A glance at Eqs. (\ref{coeff})
and (\ref{masKr})
shows that $\gamma_{\underline{0i}}$ vanishes. The
derivation of $\gamma_{\underline{00}}$ is also not difficult.
Inserting Eq. (\ref{masKr}) in Eq. (\ref{coeff}) one
sees immediately that it is given by the Taylor expansion of
Newton's potential without the first two terms. Assuming
that we are in the range of convergence of the Taylor series
we thus find
\beq \gamma_{\underline{00}} = 2 (\phi (\vec{x}) - \phi (0) -
     x^i \phi_{,i}|_0 ) \; .\eeq
Note that all acceleration-dependent contributions to $\gamma
_{\underline{\mu \nu}}$ are neglected since they are of
higher order in $h_{\mu \nu}$. To get a closed expression
for the remaining components $\gamma_{\underline{ij}}$ we
first introduce the functions
\beq v_m [f](\vec{y}) := \frac{1}{r^{m+1}} \int_0^r r^{\prime m}
     f(\vec{y}\, r^\prime /r)\, dr^\prime \eeq
where $f$ is some function of $\vec{y}$ and $r = |\vec{y}|$
is the absolute value of the three-vector $y^i$ in the usual
Cartesian sense.
One can show by induction and by using de l'Hospital's rule
that the Taylor coefficients of these functions are given by
\beq v_{m,k_1 \cdots k_n}[f] |_0 = \frac{1}{n+m+1} f_{,k_1
     \cdots k_n}|_0 \; .\label{Abl0} \eeq
After the insertion of $R_{imjn}$ from Eq. (\ref{masKr}) into
Eq. (\ref{coeff}) one can see that the resulting series
is just the Taylor expansion of
\bea \gamma_{\underline{ij}} &=& 2 \vec{x}^2 \{ v_1 [\phi_{,
     ij}] - v_2 [\phi_{,ij}]\} + 2 \delta_{ij} \{ \phi (\vec{x})
     - 2 v_0 [\phi ] + \phi (0) \} \nn \\ & &
     - 2 \Big [ x^i \{ 2 v_1 [\phi_{,j}] - v_0 [\phi_{,j}] \}
     + x^j \{ 2 v_1 [\phi_{,i}] - v_0 [\phi_{,i}] \} \Big ]
     \; . \label{ijErg} \eea
It follows that the metric in the Fermi coordinates of our
observer is given by
\bea g_{00} &=& -1 +2 (\phi (0) - \phi (\vec{x})) \nn \\
     g_{0i} &=& 0 \nn \\
     g_{ij} &=& \delta_{ij} - \gamma_{\underline{ij}} \; .
     \label{gmunuerg} \eea
This result has a clear physical interpretation. As is well
known the largest effect for objects with slow velocities comes
from the $g_{00}$ component of the metric. This effect is
identical to that of the Newtonian potential normalized to be
zero on the worldline of the observer as indicated by
Eq. (\ref{gmunuerg}).
This normalization condition has its origin in the use of the
proper time of the observer as the time coordinate. Another
choice of time would lead to a different normalization of the
potential.

Less obvious is the interpretation of the $g_{ij}$ components
which can be tested by the measurement of spatial distances.
Since the expression (\ref{ijErg}) for $\gamma_{\underline{ij}}$
involves the gradient and the second derivatives of the
potential we can infer that distances measured in
the direction of the field
gradient or of the main axes of the matrix $\phi_{,ij}$
behave differently as in other directions.
This may be more obvious for the (unnatural) case when
$\phi (\vec{y}) = \phi (r)$ is a function of the distance $r$
to the observer only. After several partial integrations
in Eq. (\ref{ijErg}) one gets
\beq \gamma_{\underline{ij}} = 2 \Big ( \frac{\delta_{ij}}{r}
     - \frac{x^i x^j}{r^3}\Big ) \int_0^r r^\prime \frac{d \phi
     }{d r^\prime} dr^\prime \eeq
where we have slightly changed the notation so that $r$ is now
identical to $|\vec{x}|$. We see
that the radial direction is indeed preferred.
\section{Conclusion}
In this paper we have shown that the linearized construction
of Fermi coordinates can be performed for arbitrary
space-time geometries, arbitrary motion of the observer,
and to all orders in the geodesic distance $s$ from the
worldline.
In particular, we have treated the case of a resting observer
in the field of a static mass distribution. This enables
us to make the correspondence between general relativity
and Newtonian gravity more precise.

One advantage of the knowledge of the metric to all orders in
the spatial geodesic distance $s$ from the worldline may be that
one can considerably enlarge the range of validity of the
Fermi coordinate system. In the usual expansion to second order
in $s$ one has, beside others, to fulfill the condition
\cite{manasse63} $|R_{\mu \nu \rho \sigma , k}| |x^k|
/ | R_{\mu \nu \rho \sigma}| \ll 1$. For a gravitational wave
with wave length $\lambda$,
for instance, the Riemann tensor is roughly proportional to
$ \exp (i k x) / \lambda^2$ so that this condition gives
$ |x^k| / \lambda \ll 1 $. For laser detectors this may be
restrictive since the wavelength is often supposed to be
in the order of 300 km. If one can calculate the whole sum
in Eq. (\ref{coeff}) the limit is much larger. For growing
distance from the worldline and certain directions the factors
$\gamma_{\underline{ij}}$ can grow roughly
\cite{vergl} like
$A (x^l/\lambda)^2$, where $A$ is the amplitude of the
gravitational wave which is usually assumed to be smaller than
about $10^{-18}$. It should be stressed that this equation of
Ref.~\cite{faraoni92}, which was first derived by Baroni
{\em et al.} \cite{baroni86}, includes the first order in
$h_{\mu \nu}$ but all orders in the geodesic distance $s$ from
the worldline. The sum of Eq.~(\ref{coeff}) was given in a
closed form in the same sense as in our
Eq.~(\ref{gmunuerg}). The only restriction is now that the
corrections to the Minkowski metric remain weak. This means
$|\gamma_{\underline{ij}}| \ll 1$, or equivalently
$ |x^l| \ll \lambda / \sqrt{A}$. We see that the knowledge
of the whole sum has enlarged the range of validity by a
factor of $1/\sqrt{A}$ which is about $10^9$ in our example.
In this case there is no problem to describe contemporary
laser detectors of gravitational waves in Fermi coordinates.
This last conclusion is not in concordance with
Ref. \cite{faraoni92}.

We would like to thank J. Audretsch for stimulating discussions
and the Studienstiftung des Deutschen
Volkes for financial support.


\begin{thebibliography}{99}
\bibitem{vergl} Ref. \cite{faraoni92}, we refer to the
 	quantities $h_{\mu \nu}^{(FNC)}$ of this publication.
\end{thebibliography}
\end{document}